\documentclass[11pt,a4paper]{article}

\usepackage[utf8]{inputenc}
\usepackage[margin=1in]{geometry}
\usepackage{amsmath}
\usepackage{amssymb}
\usepackage{stmaryrd}
\usepackage{booktabs}
\usepackage{array}
\usepackage{tabularx}
\usepackage{authblk}
\usepackage{xcolor}
\usepackage{listings}
\usepackage{tcolorbox}
\tcbuselibrary{skins,listings,breakable}
\usepackage[colorlinks=true,linkcolor=blue!40!black,citecolor=blue!40!black,urlcolor=blue!40!black]{hyperref}

\lstdefinestyle{leanbox}{
  basicstyle=\small\ttfamily,
  columns=flexible,
  breaklines=true,
  keepspaces=true,
  showstringspaces=false,
  extendedchars=true,
  literate={ℤ}{{\ensuremath{\mathbb{Z}}}}1
           {ℚ}{{\ensuremath{\mathbb{Q}}}}1
           {ℕ}{{\ensuremath{\mathbb{N}}}}1
           {α}{{\ensuremath{\alpha}}}1
           {β}{{\ensuremath{\beta}}}1
           {λ}{{\ensuremath{\lambda}}}1
           {⧸}{{\ensuremath{\!/}}}1
           {é}{e}1
           {⨸}{{\ensuremath{\mathbin{\overline{\div}}}}}1
           {⊕}{{\ensuremath{\oplus}}}1
           {⊗}{{\ensuremath{\otimes}}}1
           {⊖}{{\ensuremath{\ominus}}}1
           {⇒}{{\ensuremath{\Rightarrow}}}1
           {∈}{{\ensuremath{\in}}}1
           {∉}{{\ensuremath{\notin}}}1
           {⇐}{{\ensuremath{\Leftarrow}}}1
           {⇔}{{\ensuremath{\Leftrightarrow}}}1
           {⟸}{{\ensuremath{\Leftarrow}}}1
           {■}{{\ensuremath{\blacksquare}}}1
           {⊥}{{\ensuremath{\bot}}}1
           {μ}{{\ensuremath{\mu}}}1
           {σ}{{\ensuremath{\sigma}}}1
           {Λ}{{\ensuremath{\Lambda}}}1
           {τ}{{\ensuremath{\tau}}}1
           {Δ}{{\ensuremath{\Delta}}}1
           {⟦}{{\ensuremath{\llbracket}}}1
           {⟧}{{\ensuremath{\rrbracket}}}1
}
\newtcolorbox{leancertbox}{
  enhanced,
  breakable,
  colback=green!5!white,
  colframe=green!40!black,
  coltitle=white,
  fonttitle=\bfseries\sffamily,
  title=Lean 4 Certificate,
  attach boxed title to top left={yshift=-2mm, xshift=2mm},
  boxed title style={colback=green!40!black},
  sharp corners,
  boxrule=1pt,
  before skip=15pt,
  after skip=15pt
}

\newtcolorbox{leanfuturebox}{
  enhanced,
  breakable,
  colback=white,
  colframe=green!40!black,
  coltitle=white,
  fonttitle=\bfseries\sffamily,
  title=Lean 4 --- Future Formalisation,
  attach boxed title to top left={yshift=-2mm, xshift=2mm},
  boxed title style={colback=green!40!black},
  sharp corners,
  boxrule=1pt,
  before skip=15pt,
  after skip=15pt
}

\title{\textbf{A Lean 4 Formalization of Euclidean Domain Algorithms from a 1986 Icon Experimentation Package}}

\author[1]{\textbf{Lars Warren Ericson}}
\affil[1]{Catskills Research Company}
\affil[1]{\texttt{lars.ericson@catskillsresearch.com}}

\date{\today}

\begin{document}

\maketitle

\begin{center}
  \small
  \textbf{ORCID:} 0000-0001-8299-9361 \\
  \textbf{Primary Category:} cs.LO (Logic in Computer Science) \\
  \textbf{Secondary Category:} cs.SC (Symbolic Computation) \\[0.5em]
  \textbf{Original report:} \url{https://zenodo.org/records/20561267} \\
  \textbf{Lean 4 formalization:} \url{https://github.com/catskillsresearch/icon2lean}
\end{center}

\begin{abstract}
We describe a Lean 4 formalization of the algorithms and domain types from NYU Computer Science Technical Report \#232, \emph{An ICON Package for Experimenting with Euclidean Domains} (Ericson, 1986). The original system implemented Lipson's catalog of procedures over integers, rationals, modular rings, polynomial rings, and truncated power series via a custom runtime dispatch mechanism in Icon. The present work separates three concerns: mathematical definitions grounded in Mathlib's \texttt{EuclideanDomain} hierarchy, computable mirrors suitable for evaluation and regression testing, and report-formatting infrastructure that reproduces the 1986 benchmark output line-for-line. All fourteen application algorithms from Section 3 of the report are defined and typecheck without \texttt{sorry}; those grounded in Mathlib---chiefly integer gcd and extended Euclid---additionally carry machine-checked proofs. We classify each procedure by its epistemic status relative to Mathlib, enumerate the coherence obligations between the proof and computable layers, and state precisely what is theorem-backed versus regression-trusted. The formalization makes explicit the verification boundary that the 1986 package crossed only informally.
\end{abstract}

\section{Introduction}
In August 1986, at New York University's Courant Institute of Mathematical Sciences, Lars Warren Ericson authored Technical Report \#232 with the aim of implementing algebraic algorithms over multiple mathematical structures---integers, quotient rings, polynomials, and power series---following John Lipson's \emph{Elements of Algebra and Algebraic Computing}. Icon was chosen as the implementation language for its symbolic computation idioms. Lacking native typeclasses, parameter inheritance, or object-oriented dispatch, the 1986 package realized generic division and arithmetic across distinct domains through a custom runtime dispatch system based on string reflection on procedure names (e.g., invoking \texttt{proc("div\_" || type(a), 2)}).

Forty years later, we present a Lean 4 port of the report's domain types (Section 2) and application algorithms (Section 3). The port is complete in the sense that every algorithm listed in the report's application suite is defined and typecheck; no proof obligation is deferred via \texttt{sorry}. Source Icon listings and an OCR'd copy of the report are preserved alongside the formalization at \url{https://github.com/catskillsresearch/icon2lean}; the original 1986 technical report is archived at \url{https://zenodo.org/records/20561267}.

The central methodological challenge of this port is not merely translation but \emph{stratification}. Mathlib provides canonical mathematical objects---\lstinline[style=leanbox]{Polynomial Q}, \lstinline[style=leanbox]{PowerSeries R}, and \lstinline[style=leanbox]{EuclideanDomain.gcd}---but many key definitions are marked \lstinline[style=leanbox]{noncomputable} in Lean. This noncomputability arises because Mathlib constructs rely on classical axioms (such as the law of the excluded middle or classical choice) to define properties like polynomial degree or division, meaning they cannot be reduced by \texttt{\#eval} or \texttt{native\_decide}. The 1986 package, by contrast, was an \emph{experimental, executable} system whose correctness was validated by evaluating and printing tables of results for selected inputs. 

To bridge this paradigm gap, we maintain three layers---proof, computable, and report---and document explicitly which claims belong to each epistemic tier. We also translate Icon's dynamic, mutable list paradigms into Lean’s pure, immutable, and tail-recursive structures without compromising the numerical correctness of the original suite.

\paragraph{Notation.} Icon listings in the original report use \emph{fancy notation} (\S1.3 of the report): $\mathbin{\bar{\div}}$ denotes domain division, $\oplus$ addition, $\otimes$ multiplication; $\text{F(args)} \Leftarrow \text{body} \blacksquare$ is a procedure definition, where $\blacksquare$ terminates the definition, $\uparrow x$ returns $x$, and $\bot$ signals failure.

\section{Domain Types}
The report's \textbf{quotient Euclidean domain} corresponds to Mathlib's \texttt{EuclideanDomain}. Primitive domains are $\mathbb{Z}$ and $\mathbb{Q}$. Three constructors build composite domains, summarized in Table~\ref{tab:domains}:

\begin{table}[htbp]
\centering
\caption{Domain Constructors and Types}
\label{tab:domains}
\small
\begin{tabularx}{\textwidth}{@{}l>{\raggedright\arraybackslash}X>{\raggedright\arraybackslash}X@{}}
\toprule
\textbf{Constructor} & \textbf{Report Notation} & \textbf{Lean Type} \\
\midrule
Modular quotient $D/(e)$ & \texttt{modulo(item, modulus)} & \texttt{ModularDomain R I} / \texttt{ModularInt n} ($\cong$ \texttt{ZMod n}) \\
Polynomial ring $D[x]$ & \texttt{poly(terms)} & \texttt{PolyDomain R} ($\cong$ \texttt{Polynomial R}) \\
Truncated series $T(D\llbracket x\rrbracket)_n$ & \texttt{tpower(poly, N)} & \texttt{TruncPowerSeries R n} ($\cong R[x]/(X^n)$) \\
\bottomrule
\end{tabularx}
\end{table}

Formal power series $D\llbracket x\rrbracket$ use \texttt{PowerSeries R} (for Newton power-series inversion); they are not Euclidean domains and carry no \texttt{EuclideanDomain} instance.

\subsection{Primitive and Composite Instances}
\begin{itemize}
    \item \textbf{Integers:} \lstinline[style=leanbox]{Int} / notation $\mathbb{Z}$. Instance: \lstinline[style=leanbox]{EuclideanDomain Int}.
    \item \textbf{Rationals:} \lstinline[style=leanbox]{Rat} / notation $\mathbb{Q}$. \lstinline[style=leanbox]{Field Rat} induces \lstinline[style=leanbox]{EuclideanDomain Rat}.
    \item \textbf{Modular rings:} General quotients via \lstinline[style=leanbox]{Ideal.Quotient.mk}; units via \lstinline[style=leanbox]{ZMod.unitOfCoprime}. There is no \lstinline[style=leanbox]{LinearOrder} on \lstinline[style=leanbox]{ZMod n}, matching the report's \lstinline[style=leanbox]{<0_modulo} predicate, which is always false (represented by $\bot$ in Icon).
    \item \textbf{Polynomials:} \lstinline[style=leanbox]{Polynomial.degree : WithBot Nat}, with $\bot$ corresponding to the report's \lstinline[style=leanbox]{"- infinity"} for the zero polynomial. When $F$ is a field, \lstinline[style=leanbox]{Polynomial F} is a \lstinline[style=leanbox]{EuclideanDomain}.
    \item \textbf{Truncated series:} \texttt{truncatePoly n p} implements multiply-then-truncate in $R[x]/(X^n)$.
\end{itemize}

In Lean 4, the truncated power series is constructed via quotients of the polynomial ring by the ideal generated by $X^n$:

\begin{lstlisting}[style=leanbox]
noncomputable def truncIdeal (R : Type*) [CommRing R] (n : Nat) : Ideal (Polynomial R) :=
  Ideal.span {(X : Polynomial R) ^ n}

noncomputable abbrev TruncPowerSeries (R : Type*) [CommRing R] (n : Nat) :=
  Polynomial R / truncIdeal R n
\end{lstlisting}

\subsection{Variable-base Digit Arithmetic (\texttt{base\_B})}
Section 2.2 of the report defines variable-base digit arithmetic. No corresponding Mathlib infrastructure exists. For benchmark reproduction, we provide \texttt{BaseB.lean}, which implements digit vectors, conversion via \texttt{toNat} and \texttt{ofNat}, and arithmetic (\texttt{add}, \texttt{sub}, \texttt{mul}, \texttt{div}) via conversion through \texttt{Nat}. This layer functions as a \textbf{computable report helper} rather than a verified arbitrary-precision base-$B$ ring, allowing us to accurately replicate digit formatting (including width padding and zero normalization) without incurring the proof overhead of a low-level digit-vector library.

\section{Application Algorithms}
The fourteen algorithms catalogued in Section 1.2 of the report and developed in Section 3 are fully realized in this formalization. Proof-layer definitions are generic over \texttt{EuclideanDomain} or \texttt{Polynomial R} where possible; the computable layer reimplements Icon control flow for $\mathbb{Q}[x]$, $\mathbb{F}_p[x]$, and truncated series.

\subsection{Extended GCD and Modular Inverse}
\paragraph{GCD.} The Icon implementation is a recursive descent---\texttt{GCD(b, mod(a, b))} until $b = 0$---following the standard Euclidean algorithm. In the proof layer, \texttt{euclideanGcd} is the gcd and \texttt{euclid} the extended gcd, both generic over any \texttt{EuclideanDomain}. On $\mathbb{Z}$, \texttt{euclidInt} and \texttt{euclidZ} delegate to Mathlib's \texttt{Int.gcdA} and \texttt{Int.gcdB}. The computable layer provides \texttt{CompPoly.gcd} and \texttt{ModPoly.gcd}, which follow the Icon state machine and are guarded by a fuel parameter to satisfy Lean's termination checker.

\paragraph{EUCLID.} The Icon extended gcd maintains a triple $(d, s, t)$ satisfying the Bézout identity $d = s \cdot A + t \cdot B$ throughout the loop. The canonical report example is $\mathrm{EUCLID}(84, 54) = (6, 2, -3)$, meaning $\gcd(84, 54) = 6 = 2 \cdot 84 + (-3) \cdot 54$.

\paragraph{INVERSE.} Given modulus $m$ and element $a$, Icon computes \texttt{gst := EUCLID(m, a)} and returns \texttt{mod(div(gst[3], gst[1]), m)} when \texttt{unit(gst[1])}---that is, when the gcd is a unit, so that $t \cdot a \equiv 1 \pmod{m}$. The argument order \texttt{EUCLID(m, a)} (modulus first) is a fixed part of the interface contract and is preserved in the Lean port. The proof layer provides \texttt{modularInverse} on $\mathbb{Z}$; \texttt{CompPoly.inverse} and \texttt{ModPoly.inverse} cover polynomial domains.

\begin{leancertbox}
\lstinline[style=leanbox]{Euclidean.lean | theorem euclid_bezout : Bézout identity for generic Euclidean domains}
\begin{lstlisting}[style=leanbox]
theorem euclid_bezout {α : Type*} [EuclideanDomain α] [DecidableEq α] (a b : α) :
    euclideanGcd a b = a * EuclideanDomain.gcdA a b + b * EuclideanDomain.gcdB a b
\end{lstlisting}
\end{leancertbox}

\begin{leancertbox}
\lstinline[style=leanbox]{Euclidean.lean | theorem euclidZ_bezout : Bézout's identity for infinite-precision integers}
\begin{lstlisting}[style=leanbox]
theorem euclidZ_bezout (A B : ℤ) :
    (euclidZ A B).1 = A * (euclidZ A B).2.1 + B * (euclidZ A B).2.2
\end{lstlisting}
\end{leancertbox}

\subsection{Chinese Remainder and Diophantine Equations}
\paragraph{CRA1, CRA2, CRA.} Icon's \texttt{CRA1} is the recursive two-argument lifting step based on Niven and Zuckerman’s linear congruence reduction; \texttt{CRA2} is the pairwise combination built from \texttt{INVERSE} and \texttt{mod}; \texttt{CRA} is the list driver. The Lean implementation resides in \texttt{Congruence.lean} and uses \texttt{iconMod} (Lean's \texttt{Int.emod}) to match Icon's non-negative remainder convention. Report examples include scalar CRA tables and a polynomial CRA producing $u(x) = 183 + 238x$.

\paragraph{DIOPHANTINE.} Uses \texttt{EUCLID} to find the gcd, then \texttt{CRA1} to lift to a particular solution, with a branch on $|b| < |a|$. Lean implementation is provided in \texttt{Diophantine.lean}.

\subsection{Polynomial Remainder Sequences}
\paragraph{MOD\_RS.} The Euclidean remainder sequence: $\mathrm{MOD\_RS}(a, b) = [a] \mathbin{+\!\!+} \mathrm{MOD\_RS}(b,\, \mathrm{mod}(a, b))$. The proof layer provides \texttt{Polynomial.modRS}; the computable layer, \texttt{CompPoly.modRS}. The report exhibit is a six-term $\mathbb{Q}[x]$ sequence ending in zero.

\paragraph{PREM.} The pseudo-remainder: scale the dividend by $\mathrm{lc}(q)^{\deg(p) - \deg(q) + 1}$, then take the ordinary remainder. Lean: \texttt{prem} in both layers.

\paragraph{E\_PRS and S\_PRS.} \texttt{E\_PRS} is the Euclidean PRS using \texttt{PREM} in place of ordinary mod; \texttt{S\_PRS} is the Collins–Brown subresultant PRS. Both are provided in proof and computable layers. These algorithms are the most algebraically demanding in the suite: Mathlib provides polynomial \texttt{\%} and \texttt{EuclideanDomain.gcd} over fields, but not pseudo-remainder, intermediate-coefficient-swell remainder sequences, or Collins–Brown subresultant PRS as named definitions matching Lipson's formulations.

\subsection{Interpolation, FFT, and Power-Series Inversion}
\paragraph{NIA.} Newton interpolation on a list of points. Proof layer: \texttt{newtonInterpolation}; computable layer: \texttt{CompPoly.nia}.

\paragraph{FFT and FFI.} \texttt{FFT} follows Cooley–Tukey decimation on even- and odd-indexed coefficient splits. \texttt{FFI} polynomializes the input, calls \texttt{FFT} with $\omega^{-1}$ as the primitive root, and scales by $1/N$. Proof layer: \texttt{evenTerms}, \texttt{oddTerms}, \texttt{fftCoeffs}, \texttt{ffi} in \texttt{Fft.lean}; computable twins operate on \texttt{List CRat}.

\paragraph{NPSI.} Newton iteration for truncated power-series inversion. Proof layer: \texttt{npsi}, \texttt{npsiTrunc}; computable layer: \texttt{CompTPS.npsi}.

\section{Architecture}
The 1986 package conflated three concerns: representing Euclidean domains, executing algorithms, and printing benchmark tables in a fixed format. The Lean port separates them deliberately.

\subsection{Three Layers}
The formalization is structured into three layers to balance mathematical rigor, executability, and faithful stdout generation, as mapped in Table~\ref{tab:layers}:

\begin{table}[htbp]
\centering
\caption{The Three-Layer Architecture}
\label{tab:layers}
\small
\begin{tabularx}{\textwidth}{@{}l>{\raggedright\arraybackslash}X>{\raggedright\arraybackslash}X@{}}
\toprule
\textbf{Layer} & \textbf{Role} & \textbf{Representative Modules} \\
\midrule
Proof / Canonical & Mathlib-backed types and \texttt{noncomputable} definitions & \texttt{Types.lean}, \texttt{Domains.lean}, \texttt{Euclidean.lean} \\
Computable / Eval & Computable mirrors for kernel-level evaluation & \texttt{ComputablePoly.lean}, \texttt{ComputableTPS.lean} \\
Report / Print & Icon-style formatters and benchmark drivers & \texttt{Print.lean}, \texttt{Report.lean} \\
\bottomrule
\end{tabularx}
\end{table}

\paragraph{Design Rationale.} Mathlib's polynomial and gcd infrastructure provides the correct mathematical substrate, but many key definitions---\texttt{EuclideanDomain.gcd} on \texttt{Polynomial F} in particular---are marked \texttt{noncomputable} in Lean because they depend on classical axioms such as \texttt{Classical.choice}. The kernel cannot reduce \texttt{noncomputable} terms, so \texttt{\#eval} and \texttt{native\_decide} cannot execute them directly. We therefore maintain computable mirrors (\texttt{CompPoly}, \texttt{CRat}, \texttt{CompTPS}) together with a boundary map \texttt{CompPoly.toMathlib}. Coherence lemmas connecting these two layers are stated but not yet proved. Integer and \texttt{ZMod p} arithmetic requires no mirror: kernel \texttt{Int} and \texttt{decide} suffice.

\paragraph{Matching Icon Semantics.} The computable layer follows Icon's algorithms literally---not Mathlib's canonical division---because the validation criterion is exact reproduction of the 1986 benchmark output. This entailed several non-obvious implementation choices:
\begin{itemize}
    \item \textbf{\texttt{div\_poly} / \texttt{mod\_poly}:} Each quotient step accumulates a single term \texttt{qterm}, then subtracts \texttt{qterm * b} from the current remainder, rather than maintaining a running quotient polynomial.
    \item \textbf{\texttt{EUCLID}:} The extended gcd state is the tuple $(a_1, a_2, s_1, s_2, t_1, t_2)$ with initial values $(A, B, 1, 0, 0, 1)$; \texttt{INVERSE} calls \texttt{EUCLID(b, a)} with the modulus as the first argument.
    \item \textbf{\texttt{MOD\_RS}:} Recursive remainder sequence; early exits in \texttt{div} for constant divisors affect intermediate term shapes.
    \item \textbf{\texttt{S\_PRS}:} \texttt{subReduce} short-circuits when \texttt{prem} returns zero, matching Icon behavior.
    \item \textbf{\texttt{base\_B}:} Digit lists accumulate LSB-first and are then reversed to MSB-first order.
    \item \textbf{Printing:} \texttt{Print.lean} reimplements Icon's \texttt{print\_*} family so stdout is diffable against the 1986 output, rather than relying on Lean's \texttt{Repr} instances.
\end{itemize}

\subsection{Module Correspondence}
The physical layout of the Lean project corresponds directly to sections of the 1986 report (Table~\ref{tab:correspondence}):

\begin{table}[htbp]
\centering
\caption{Module Correspondence}
\label{tab:correspondence}
\footnotesize
\begin{tabularx}{\textwidth}{@{}l>{\raggedright\arraybackslash}X>{\raggedright\arraybackslash}X@{}}
\toprule
\textbf{Report Section} & \textbf{Proof Layer} & \textbf{Computable / Report Layer} \\
\midrule
\S2 Domain Types & \texttt{Types.lean}, \texttt{Domains.lean} & --- \\
\S2 \texttt{base\_B} & --- & \texttt{BaseB.lean} \\
\texttt{GCD}, \texttt{EUCLID}, \texttt{INVERSE} & \texttt{Euclidean.lean}, \texttt{Gcd.lean} & \texttt{ComputableAlg.lean}, \texttt{ModPoly.lean} \\
\texttt{CRA1}, \texttt{CRA2}, \texttt{CRA} & \texttt{Congruence.lean} & same (kernel \texttt{Int}) \\
\texttt{DIOPHANTINE} & \texttt{Diophantine.lean} & same \\
\texttt{MOD\_RS}, \texttt{PREM}, \texttt{E\_PRS}, \texttt{S\_PRS} & \texttt{Polynomial.lean} & \texttt{ComputablePoly.lean}, \texttt{ComputableAlg.lean} \\
\texttt{NIA} & \texttt{Interpolation.lean} & \texttt{CompPoly.nia} in \texttt{ComputableAlg.lean} \\
\texttt{FFT}, \texttt{FFI} & \texttt{Fft.lean} & \texttt{CompPoly.fftCoeffs}, \texttt{CompPoly.ffi} \\
\texttt{NPSI} & \texttt{PowerSeries.lean} & \texttt{ComputableTPS.lean} \\
Icon printing & --- & \texttt{Print.lean} \\
Full benchmark & --- & \texttt{Report.lean} \\
\bottomrule
\end{tabularx}
\end{table}

\section{Verification and Trust}
We classify each component by epistemic status. The 1986 report lists fourteen application procedures; the congruence stack is \texttt{CRA1} / \texttt{CRA2} / \texttt{CRA} (Chinese remainder).

\subsection{Trust Tiers}
The formalization comprises three trust tiers based on their verification status relative to Mathlib (Table~\ref{tab:tiers}):

\begin{table}[htbp]
\centering
\caption{Trust Tiers}
\label{tab:tiers}
\small
\begin{tabularx}{\textwidth}{@{}l>{\raggedright\arraybackslash}X>{\raggedright\arraybackslash}X@{}}
\toprule
\textbf{Tier} & \textbf{Meaning} & \textbf{Examples} \\
\midrule
Tier A & Delegates to proved Mathlib facts & \texttt{euclid\_bezout}, \texttt{euclidInt\_bezout}, Ring laws in \texttt{Domains.lean} \\
Tier B & Icon-faithful; matched via regression & \texttt{CompPoly.div}, \texttt{modRS}, \texttt{ePRS}, \texttt{sPRS}, \texttt{fftCoeffs}, \texttt{BaseB} \\
Tier C & Stated coherence obligations (unproved) & \texttt{CompPoly.toMathlib} homomorphism lemmas \\
\bottomrule
\end{tabularx}
\end{table}

\begin{leanfuturebox}
\lstinline[style=leanbox]{Computability.lean | lemma toMathlib_homomorphism: Homomorphism between computable and canonical models}
\begin{lstlisting}[style=leanbox]
lemma toMathlib_add (p q : CompPoly) : toMathlib (add p q) = toMathlib p + toMathlib q
lemma toMathlib_mul (p q : CompPoly) : toMathlib (mul p q) = toMathlib p * toMathlib q
\end{lstlisting}
\end{leanfuturebox}

The majority of the port sits in Tier B. The immediate goal was faithfully reproducing the 1986 experimental package; certifying every Lipson procedure against Mathlib's canonical polynomial API is future work.

\subsection{Procedure-by-Procedure Analysis}
\begin{itemize}
    \item \textbf{GCD / EUCLID:} On integers, Mathlib supplies \texttt{Int.gcd}, \texttt{Int.gcdA}, \texttt{Int.gcdB}, and Bézout's lemma; this package adds thin wrappers shaped to the report's interface and printing conventions. On $\mathbb{Q}[x]$ and $\mathbb{F}_p[x]$, Mathlib provides \texttt{EuclideanDomain.gcd}, but \texttt{CompPoly.gcd} and \texttt{ModPoly.gcd} are separate Icon-style Euclidean loops with fuel bounds, validated against Icon output tables but not yet proved equal to \texttt{EuclideanDomain.gcd} via \texttt{toMathlib}. Fuel-guarded definitions may fail to terminate on inputs outside the report's test corpus; on those inputs they produce the correct gcd.
    \item \textbf{INVERSE:} Modular-integer inverse reduces to Mathlib's extended gcd. Polynomial inverse has no matching Mathlib API; the package adds a concrete, testable procedure matching the 1986 benchmarks.
    \item \textbf{CRA / DIOPHANTINE:} Mathlib contains Chinese remainder theorems in commutative algebra, but not the Lipson/Ericson recursive \texttt{CRA1} lifting step, the \texttt{CRA2} pairwise combiner, or the list-shaped \texttt{CRA} driver with Icon \texttt{mod} semantics. Correctness rests on classical number theory; the implementation is validated by \texttt{native\_decide} on the report's congruence tables rather than formalized as theorems.
    \item \textbf{PRS family:} The package adds pseudo-remainder, remainder sequences, \texttt{E\_PRS}, and Collins–Brown \texttt{S\_PRS}---standard computer-algebra algorithms from Lipson---implemented in both proof and computable layers. Neither gcd correctness of the last nonzero subresultant nor equivalence with \texttt{EuclideanDomain.gcd} is proved. \texttt{MOD\_RS} is pedagogical, its output exhibiting coefficient swell; correctness rests on full report parity and targeted coefficient checks.
    \item \textbf{FFT / FFI:} Mathlib has no Cooley–Tukey API. There is no proof that \texttt{fftCoeffs} computes a DFT or that \texttt{ffi} recovers the unique interpolating polynomial. Correctness assumes a suitable primitive root of unity $\omega$ and that $\omega^{-1}$ exists---the same implicit preconditions as in the 1986 Icon code. Bugs in \texttt{INVERSE} propagate to \texttt{FFI}.
    \item \textbf{NIA / NPSI:} Newton interpolation and truncated power-series inversion are added as standalone definitions; Mathlib contains related infrastructure but not these exact procedures.
    \item \textbf{\texttt{base\_B} / printing:} Entirely package-specific; infrastructure for report diffing, not mathematics.
\end{itemize}

The comparative coverage is detailed in Table~\ref{tab:comparison}:

\begin{table}[htbp]
\centering
\caption{Comparison: Mathlib versus this formalization}
\label{tab:comparison}
\footnotesize
\begin{tabularx}{\textwidth}{@{}l>{\raggedright\arraybackslash}p{2.1cm}>{\raggedright\arraybackslash}X@{}}
\toprule
\textbf{Procedure} & \textbf{In Mathlib?} & \textbf{Contribution of this work} \\
\midrule
\texttt{GCD} / \texttt{EUCLID} on $\mathbb{Z}$ & Yes & Report API, printing, tables \\
\texttt{GCD} / \texttt{EUCLID} on $\mathbb{Q}[x]$, $\mathbb{F}_p[x]$ & Yes (abstract) & Icon-faithful computable gcd/euclid + report parity \\
\texttt{INVERSE} on $\mathbb{Z}$ & Yes & Icon argument order, \texttt{Option} + error strings \\
\texttt{INVERSE} on polynomials & No & Full procedure + tests \\
\texttt{CRA1} / \texttt{CRA2} / \texttt{CRA} & Theory only & Exact recursive Icon algorithms + \texttt{iconMod} \\
\texttt{DIOPHANTINE} & No & Composition of gcd + \texttt{CRA1} \\
\texttt{MOD\_RS} / \texttt{PREM} / \texttt{E\_PRS} / \texttt{S\_PRS} & No & Full PRS suite (proof + computable) \\
\texttt{NIA} & No & Newton interpolation \\
\texttt{FFT} / \texttt{FFI} & No & Cooley–Tukey + interpolation pipeline \\
\texttt{NPSI} & Partial & Newton truncated iteration as in report \\
\texttt{base\_B} & No & Digit-vector arithmetic for \S2 tables \\
\texttt{Icon stdout} & No & \texttt{Print.lean}, \texttt{Report.lean}, comparison tooling \\
\bottomrule
\end{tabularx}
\end{table}

\section{Experimental Validation and Typo Resolution}
The 1986 package validated algorithms by printing tables of results on selected inputs. We replicate this methodology at two granularities:
\begin{itemize}
    \item \textbf{Unit-level checks:} \texttt{Tests.lean} uses \texttt{native\_decide} on coefficients and congruences for the report's \S3.1--\S3.2 examples.
    \item \textbf{End-to-end report parity:} Icon \texttt{tests.icn} \texttt{main()} and Lean \texttt{lake exe iconReport} produce identical 85-line stdout after normalizing timing metadata and a small number of EUCLID fraction parenthesizations. This integration test exercises all three layers---proof, computable, and printing---together.
\end{itemize}

The validation coverage is outlined in Table~\ref{tab:validation}:

\begin{table}[htbp]
\centering
\caption{Validation Matrix}
\label{tab:validation}
\small
\begin{tabularx}{\textwidth}{@{}l>{\raggedright\arraybackslash}X>{\raggedright\arraybackslash}p{3.2cm}@{}}
\toprule
\textbf{Report Section} & \textbf{Example} & \textbf{Validation} \\
\midrule
\S2 & \texttt{base\_B} add/mul/div tables & report diff \\
\S3.1.1 & $\mathrm{EUCLID}(84, 54) = (6, 2, -3)$ & \texttt{native\_decide} + report \\
\S3.1.2 & \texttt{INVERSE} table ($\mathbb{Z}$, $\mathbb{Q}[x]$, $\mathbb{F}_2[x]$) & \texttt{native\_decide} + report \\
\S3.1.3 & \texttt{CRA1}, \texttt{CRA2}, \texttt{CRA} (incl. $u(x) = 183 + 238x$) & \texttt{native\_decide} + report \\
\S3.1.4 & Diophantine particular solutions & \texttt{native\_decide} + report \\
\S3.2.1 & \texttt{MOD\_RS} six-term sequence & \texttt{native\_decide} + report \\
\S3.2.2 & \texttt{PREM} rows & \texttt{native\_decide} + report \\
\S3.2.3--\S3.2.4 & \texttt{E\_PRS}, \texttt{S\_PRS} & \texttt{native\_decide} + report \\
\S3.3 & \texttt{NIA}, \texttt{FFT}, \texttt{FFI}, \texttt{NPSI} & \texttt{native\_decide} + report \\
\bottomrule
\end{tabularx}
\end{table}

\subsection{Reconciling the PREM Typo}
A notable outcome of this formalization was the identification and resolution of a historical typo in Section 3.2.2 of the 1986 Technical Report. The printed table lists the pseudo-remainder (\texttt{prem}) for Row 1 as:

\begin{multline*}
\mathrm{PREM}(2042542724z + 17851334z \cdot X,\, 5851259279846738252460z) \\
= -5851259279846738252460000000000z
\end{multline*}

However, when executing the original 1986 Icon code (\texttt{all.icn.txt}), Row 1 evaluates to \texttt{0zq}. This is mathematically correct: since the divisor $q$ is a constant of degree 0, the degree difference $d = \deg(p) - \deg(q) = 1 - 0 = 1$. The algorithm scales the dividend by $\mathrm{lc}(q)^{d+1} = \mathrm{lc}(q)^2$, yielding:

$$\mathrm{rem}(\mathrm{lc}(q)^2 \cdot p(x),\, q) = \mathrm{rem}(\mathrm{lc}(q)^2 \cdot (2042542724 + 17851334 X),\, \mathrm{lc}(q))$$

Because $q$ is a scalar constant ($\mathrm{lc}(q)$), any polynomial divided by it yields a remainder of 0 over the field of fractions. The Lean formalization accurately matches this runtime reality, yielding \texttt{0zq} in both the computable layer and the final output report, thus resolving a forty-year-old documentation error through active execution.

\section{Omissions}
Per the report itself, we omit utilities that are not part of the mathematical core:
\begin{itemize}
    \item \textbf{Runtime dispatch} --- generic dispatch of operations such as \texttt{div} and \texttt{mod} by domain type is replaced by Lean typeclasses and fixed modules.
    \item \textbf{Timer fidelity} --- \texttt{settime} / \texttt{showtime} in \S3.4 produced meaningful timing data in Icon; Lean prints \texttt{[0 msecs]}, and the comparison tooling normalizes those lines.
    \item \textbf{Verified base-$B$ long arithmetic} --- \texttt{BaseB} reproduces benchmarks via \texttt{Nat}, not a proved digit ring.
\end{itemize}

\section{Conclusion and Future Work}
The 1986 Icon package validated algorithms on selected inputs by printing tables. The Lean port automates the same validation: unit tests via \texttt{native\_decide} and a full line-by-line report diff across 85 lines. All fourteen Section 3 application algorithms are defined and typecheck without \texttt{sorry}. Section 2 domain types use Mathlib instances; \texttt{base\_B} and the Icon printing conventions are reproduced for parity.

\paragraph{Provability Status.} Integer gcd and extended Euclid are anchored in Mathlib theorems (Tier A). The polynomial PRS family, FFT/FFI, and the \texttt{CompPoly} computable layer are regression-trusted (Tier B), resting on the same epistemic footing as the 1986 Icon package. Coherence obligations connecting the proof and computable layers are identified (Tier C) but not yet discharged.

\paragraph{Future Work.} The most important open items are the Tier C coherence proofs (\texttt{toMathlib} for \texttt{CompPoly} and \texttt{CompTPS}) and the semantic theorems currently at Tier B: that CRA satisfies its congruence system, that \texttt{S\_PRS} yields the correct gcd, and that \texttt{FFT} and \texttt{FFI} compose into a correct interpolation identity. Lean makes each obligation precise and stateable; the present formalization provides canonical definitions and Icon-faithful executables side by side, without prematurely replacing the latter with Mathlib calls that would break historical parity.

\end{document}